# Visual Attention Network for Low Dose CT

Wenchao Du, Hu Chen*, Peixi Liao, Hongyu Yang, Ge Wang, *Fellow, IEEE,* and Yi Zhang*, *Senior Member, IEEE*

***Abstract*—Noise and artifacts are intrinsic to low dose CT (LDCT) data acquisition, and will significantly affect the imaging performance. Perfect noise removal and image restoration is intractable in the context of LDCT due to the statistical and technical uncertainties. In this paper, we apply the generative adversarial network (GAN) framework with a visual attention mechanism to deal with this problem in a data-driven/machine learning fashion. Our main idea is to inject visual attention knowledge into the learning process of GAN to provide a powerful prior of the noise distribution. By doing this, both the generator and discriminator networks are empowered with visual attention information so they will not only pay special attention to noisy regions and surrounding structures but also explicitly assess the local consistency of the recovered regions. Our experiments qualitatively and quantitatively demonstrate the effectiveness of the proposed method with clinic CT images.**

## I. Introduction

RECENTLY, improving image quality of low-dose CT (LDCT) scans has been a hot topic. There were a large number of papers on this topic. The early methods [1, 2, 3] were based on filtering in the sinogram where the noise property is statistically known. However, any structure distortions in the sinogram domain might lead to disturbing artifacts and resolution loss in the image domain. On the other hand, iterative reconstruction methods [4, 5] can mitigate this problem to a certain degree by optimizing an objective function, which contains handcrafted prior terms, such as roughness penalty and nuclear norm. The involvement of extensive computational cost and the difficulty in designing proper regularization terms and relaxation coefficients present challenges for practical using. Recently, deep learning (DL) has been introduced into the iterative methods to solve these problems [6, 7, 8], but access to the raw data from the commercial scanners may be an obstacle for most users.

As a competitive alternative, post-processing methods [9, 10, 11, 12, 13] need not to access the raw data and are more convenient to be deployed into current CT systems. Generally, the DL-based methods [14, 15, 16] attempt to learn a nonlinear mapping from a LDCT image/sinogram to an improved counterpart by minimizing the mean squared error (MSE) loss function, which could, however, over-smooth structural details. In this paper, to alleviate this problem, inspired by the human visual perception [17] and attention model [18] for image recognition, we propose to incorporate the visual attention mechanism into the generative adversarial network (GAN) [19] framework to improve the performance of noise reduction and detail preservation. The introduction of visual attention aims to approximately locate regions contaminated by noise and guides the generator and discriminator to pay more attention on these regions and their surrounding structures. The rest of the paper is organized as follows. Section 2 introduces the proposed method. Section 3 presents experimental results. Finally, the conclusion is drawn in Section 4.

## II. Method

### A. CT Image Denoising Formation

A LDCT image can be modeled as a summation of a normal-dose CT (NDCT) image and noise:

$$x = F(y) + \varepsilon \tag{1}$$

where $x \in \mathbb{R}$ denotes a LDCT image and $y \in \mathbb{R}$ is the corresponding NDCT image. $F$ represents the degrading process caused by the noise generated from Poisson data and detector electron fluctuations. $\varepsilon$ denotes extra noise and other unmodeled factors.

The DL-based methods model the denoising procedure by learning a nonlinear mapping function $F'$ and make $F' \approx F^{-1}$ to retrieve the feasible image $\hat{y}$ . The formation is defined as:

$$F'(x) = \hat{y} \approx y \tag{2}$$

Due to the lack of prior knowledge of noise distribution, directly learning the mapping function $F'$ is difficult. To conquer this obstacle, we introduce extra visual attention map $M$ as a prior of the noise distribution to help learning $F'$ , which is implemented with a recurrent neural network. To characterize the impacts of noise on different regions in an image, Eq.(2) could be refined as:

$$F'(x \oplus M) = \hat{y} \tag{3}$$

where $M$ is a binary 2D mask, $M(i, j) = 1$ means the pixel

This work was supported in part by the National Natural Science Foundation of China under grants 616171312 and 61871277, and the Guangdong Provincial Key Laboratory of Medical Image Processing (2017B030314133).

W. Du, H. Chen, H. Yang and Y. Zhang are with the College of Computer Science, Sichuan University, Chengdu 610065, China. Y. Zhang is also with the Guangdong Provincial Key Laboratory of Medical Image Processing, Southern Medical University, Guangzhou 510515, China. Email: wenchaodu.scu@gmail.com; huchen@scu.edu.cn; yanghongyu@scu.edu.cn; yzhang@scu.edu.cn.

P. Liao is with Department of Scientific Research and Education, The Sixth People's Hospital of Chengdu, Chengdu 610065, China. Email: universe6527@163.com.

G. Wang is with Department of Biomedical Engineering, Rensselaer Polytechnic Institute, Troy, NY 12180 USA. Email: wangg6@rpi.edu.

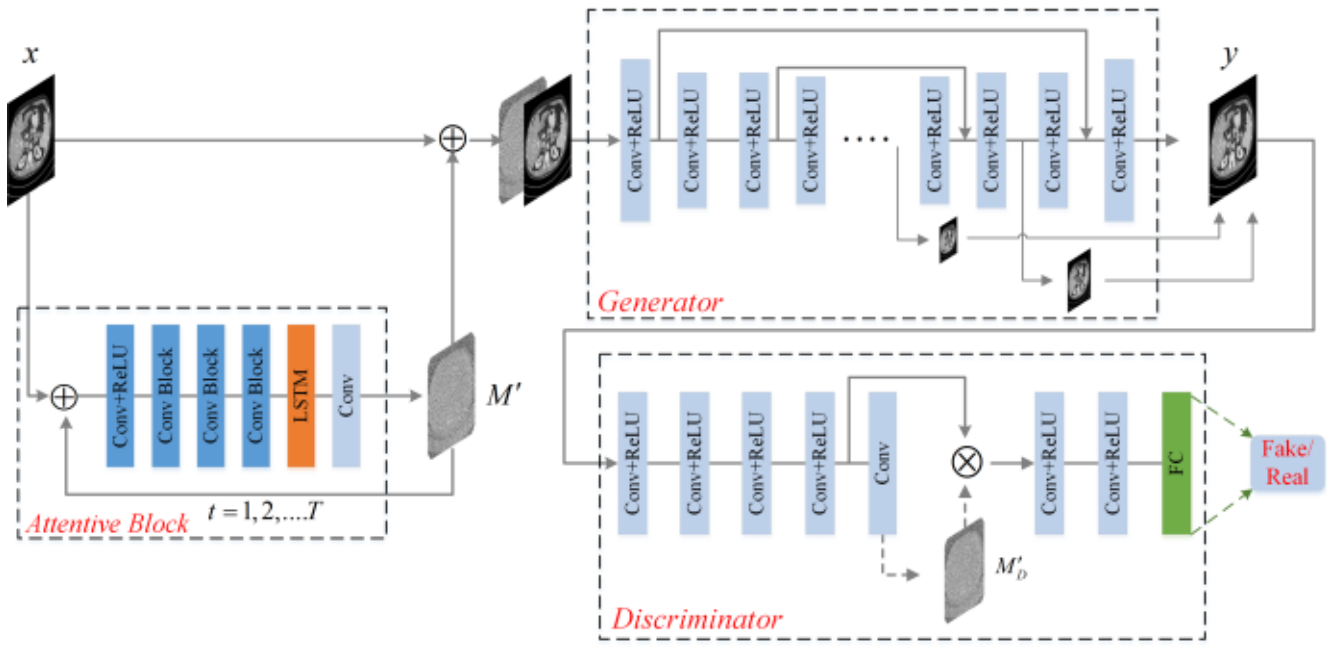

Fig. 1. Pipeline of our visual attentive network based on GAN, which includes a generator and a discriminator. An attention block is used to generate a map of noise distribution, which would be injected into generator and discriminator to guide noise removal and structural detail restoration.

$x_{(i,j)}$ is contaminated by noise, and otherwise noiseless; and the operator $\oplus$ represents the channel-wise concatenation operation. Furthermore, in order to recover the finer structural details of denoising CT image, as shown in Fig. 1, we embed our model into the framework of GAN. The attention map $M$ is injected into the generator and discriminator to guide the noise removal and structural detail preservation. The loss function of GAN is defined as:

$$\min_G \max_D \mathbb{E} \quad (D(y))] \\ + \mathbb{E} \quad (1 - D(G(x)))] \tag{4}$$

where $G$ stands for the generator and $D$ denotes the discriminator. $x$ and $y$ are the samples from the LDCT image distribution $P_L$ and NDCT distribution $P_R$ respectively.

## B. Visual Attention oriented GAN (VAGAN)

### 1) Generator with Visual Attention

Our generator aims to remove noise effectively while preserve the finer structural details from the given LDCT images. Based on these two targets, the generator consists of two components: an attentive block and a denoising generator. The attentive block aims to locate the contaminated pixels and extract surrounding structure information in the LDCT image. The output of the attentive block is the estimated knowledge about the noise distribution, which would help the generator and discriminator to focus on the noisy regions.

To achieve this goal, we first need to define the attention map $M$, which actually is a binary mask, indicating which regions in LDCT image are disturbed by noise. Note that, during the inference stage, we need estimate $M'$ from LDCT without corresponding NDCT. As the result, $M$ would serve as the prior information to guide the network to learn $M'$ directly. In the LDCT images, the noise may increase or decrease the gray values of NDCT image simultaneously in different pixel positions and the noise levels vary in different tissues. Based on the fact, residual map, which has implicitly guided the design of the network structure in some typical works [20, 21, 22], is used to approximately estimate the noise in LDCT images and we compare the gray values in the residual map with the mean of the whole residual map, which is actually an adaptive thresholding method. $M$ could be defined as follows:

$$M_{(i,j)} = \begin{cases} 1 & x'_{(i,j)} > \bar{x} \\ 0 & x'_{(i,j)} \le \bar{x} \end{cases} \tag{5}$$

where $\bar{x} = mean(x')$ and $x' = |y - x|$.

Considering that directly learning $M'$ is difficult due to the complex noise distribution, as shown in the left part of Fig. 1, we used a recurrent neural network to learn $M'$ progressively. In each time step $t$, we concatenate the input and generated $M'_t$ and feed them into the next attentive block. The proposed attentive block contains one convolutional layer, three convolutional blocks, a LSTM unit and one convolutional layer. The convolutional block is consists of two convolutional layers with residual connection. The part before the LSTM unit is used to extract the features from the LDCT images and the left components are employed to generate $M'$. Different from $M$, the estimated $M'$ is a matrix with continuous values ranging from 0 to 1, which means the greater the value is, the noisier the corresponding region is.

Since the binary mask $M$ has been acquired, it serves as a prior to supervise the generation of the estimated attention map $M'$. Therefore, a mean squired error (MSE) loss function is defined between $M'$ and $M$ at each time step $t$ and we applied $T$ steps to form the loss function as:

$$\mathcal{L} \quad \sum_{t=1}^{T} \quad \mathcal{L} \tag{6}$$

where $M'_t$ is the attention map produced by the attentive block at a time step $t$. To balance the tradeoff between the performance and computational cost, we experimentally set $T$ and the weights $\alpha$ to 4 and $[0.125, 0.25, 0.5, 1.0]$ respectively.

The residual encoder-decoder (RED) architecture has been proven effective in image denoising, deblurring or super-resolution [20, 21]. Thus, we used a similar architecture as the backbone of generator, which has 14 convolution blocks (Conv+ReLU) and skip connection is added to prevent blurring effect. A hybrid loss function is defined as follows:

$$\mathcal{L} \quad \mathcal{L} \quad \mathcal{L} \quad \mathcal{L} \quad \mathcal{L} \tag{7}$$

where $L_{GAN}(G(x)) = \log(1 - D(G(x)))$, $L_{MS}$ is the multi-scale loss used to capture the additional structural and contextual information on different scales, which is defined as

$$L_{MS}(\{x\},\{y\}) = \sum_{i=1}^{M} \lambda_i L_{MSE}(x_i, y_i) \tag{8}$$

where $x_i$ denotes the $i_{th}$ output extracted from the corresponding deconvolutional layer, and $y_i$ is the ground truth on the same scale $x_i$. $\lambda_i$ is the weight for the $i_{th}$ scale, which gradually increases as the scale increases. Specifically, the outputs of $1^{st}$, $3^{rd}$ and $5^{th}$ layers are extracted and the corresponding weights are set to 0.25, 0.5 and 1.0 in this study. $L_{Pcep}$ is the perceptual loss implemented by a pretrained VGG model (e.g., VGG19 [23] pretrained on ImageNet dataset) and usually employed to measure the similarity in the feature space:

$$L_{Pcep}(G(x), y) = \mathcal{L} \quad (x)), V(y)) \tag{9}$$

where $V(\cdot)$ denotes the pretrained VGG model.

### 2) Attentive Discriminator

The GAN-based methods for image restoration aim to recover the finer structural details. Recently, some novel approaches [24, 25, 26] have been introduced to enhance the ability of the discriminator, such as patch GAN [25] and multi-scale GAN [26], which simultaneously adopt global and local image-content consistency in the discriminative part from different scales. However, these discriminators focus on the whole image/patch to check if it is consist, which is not suitable for LDCT denoising due to the nonstationary distribution of the noise in image domain. Hence, instead of discriminating the whole image-content directly, proposed discriminator, named attentive discriminator, attempt to learn a latent noise distribution to distinguish the generated image and real NDCT image.

TABLE I
SUMMARY OF TRAINED NETWORKS: THE LOSS FUNCTIONS AND TRAINED NETWORKS

| Methods | Loss Functions |
|---|---|
| $G_{PSNR}$ | $\min_{G_{PSNR}}(\mathcal{L}$ |
| $G_{A-PSNR}$ | $\min_{G_{A-PSNR}}(\mathcal{L} \quad \mathcal{L}$ |
| $G_{GAN}$ | $\min_{G}\max_{D}\mathcal{L} \quad \mathcal{L} \quad \mathcal{L}$ |
| $G_{(A+D)-GAN}$ | $\min_{G}\max_{D}\mathcal{L} \quad \mathcal{L} \quad \mathcal{L} \quad \mathcal{L}$ |
| $VAGAN$ | $\min_{G}\max_{D}\mathcal{L} \quad \mathcal{L} \quad \mathcal{L} \quad \mathcal{L} \quad \mathcal{L}$ |

The attentive discriminator is designed to perform region-specific validation, as shown in the bottom-right of Fig.1, which is particularly useful for fine structure restoration. We first employ the attention map $M'_T$ from the attentive block at last time step $T$ in generator as the supervised information to guide discriminator from interior layers to generate a attention map $M'_D$. The original features from interior layers multiply with $M'_D$ and the newly generated feature map is feed into the following layers to decide whether the input image is artificial or real. The loss function of discriminator includes two parts, a GAN loss $\mathcal{L}$ and a attention loss $\mathcal{L}$, which could be expressed as:

$$\mathcal{L} \quad y, M'_T) = -\log(D(y)) - \log(1 - D(G(x))) + \gamma\mathcal{L} \quad y, M'_T) \tag{10}$$

where $\mathcal{L}$ is the loss between the features extracted from interior layers of discriminator and $M'_T$:

$$\mathcal{L} \quad \mathcal{L} \quad , M'_T) + \mathcal{L} \quad (y), \mathbf{0}) \tag{11}$$

where $D_{map}$ represents the procedure of generating the attention map in discriminator. In our study, $\gamma$ was empirically set to 0.05, $y$ is a randomly selected NDCT sample and $\mathbf{0}$ denotes a map only containing 0. Thus, the second term of Eq. (11) implies that for $y$, there is no region necessary to focus on. Specifically, in this work, the proposed attentive discriminator has eight convolution layers (kernel size $3\times3$) followed by a fully connected layer with 512 neurons and a sigmoid activation function.

## III. Experiments

### A. Dataset

In this study, the Mayo public clinical CT dataset was used, which was prepared for "the 2016 NIH-AAPM-Mayo Clinic

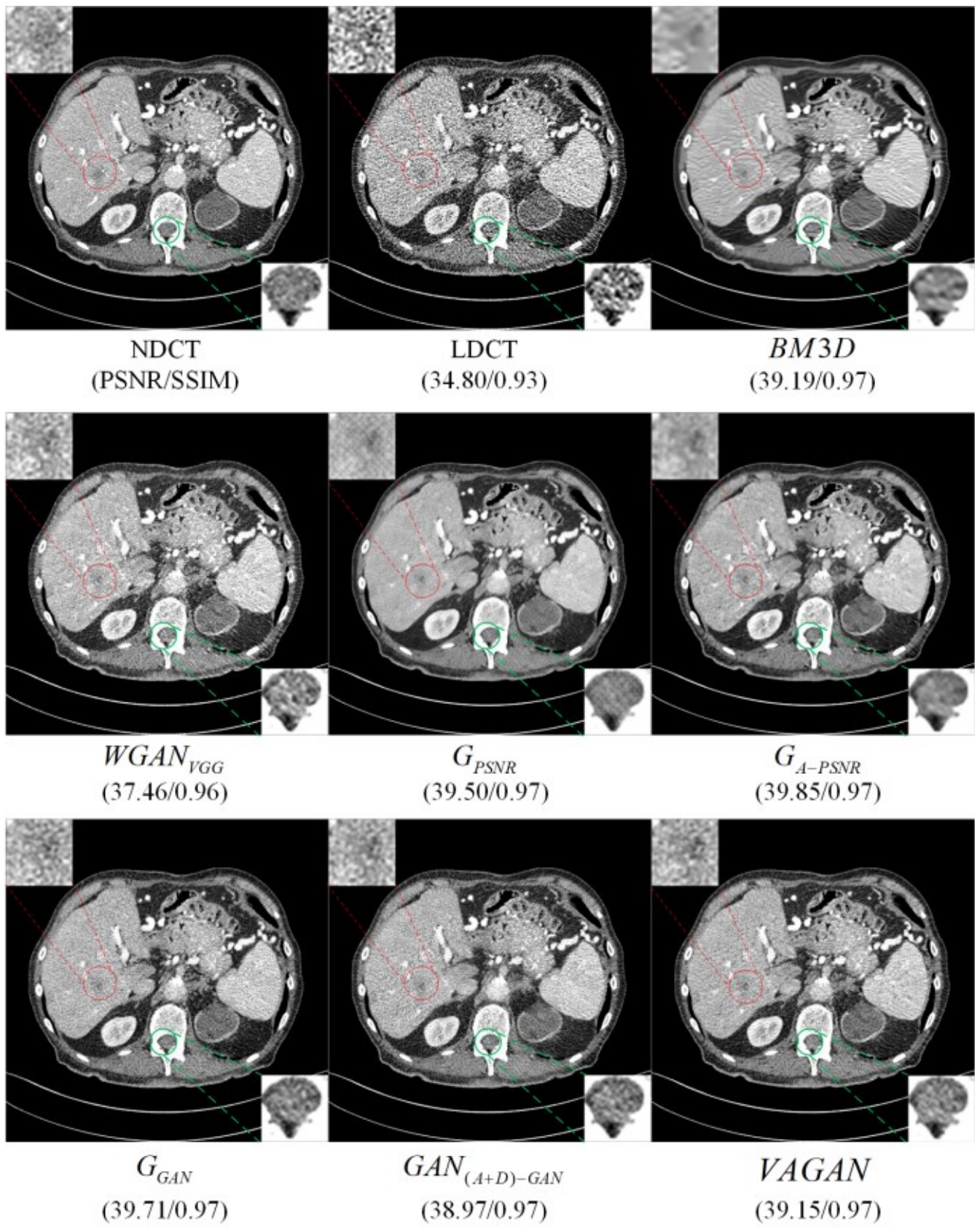

Fig. 2. Results from a transverse abdomen CT image. The display window is [-160, 240] HU.

Low Dose CT Ground Challenge" to evaluate competing LDCT image reconstruction algorithms [27]. The dataset consists of 5936 images from 10 anonymous patients' NDCT images and corresponding simulated LDCT (1/4 dose) images after realistic noise insertion. In our experiments, we randomly extracted 140,000 pairs of image patches from 4,000 slices as our training set. The patch size was 64×64. Also, we extracted 7,744 pairs of patches from another 1,936 images for testing. The peak-to-noise (PSNR) and structural similarity (SSIM) [28] are used as quantitative indexes to evaluate the performance of proposed method. In addition, the Fréchet Inception Distance (FID) [29] is introduced as perception index to quantify the visual similarity for generated denoising images.

### B. Training Details

In our experiments, the initial attention map was set to 0.5. The networks were optimized using the Adam [30] algorithm. The batch size was set to 32. The training process is divided into two stages. First, we trained a PSNR-oriented model with loss $\mathcal{L} \quad \mathcal{L} \quad \mathcal{L}$. The learning rate was initialized as 0.0002, and the total epoch numbers were set to 20. We employed a pretrained model as an initialization for generator with learning rate 0.0001 to train GAN, which mainly help generator avoid undesired local optima and obtain visually pleasing results. The method was implemented in Pytorch [31]. An NVIDIA Titan V GPU was used and the total training time is about ten hours.

To verify the effectiveness of the proposed method, we compared each network components with different loss function. Firstly, for generator, an PNSR-oriented generator network without attention map $G_{PSNR}$ and a generator with attention map $G_{A-PSNR}$ , a generator and discriminator without attention maps $G_{GAN}$ , a generator with attention map and discriminator without attention map $G_{(A+D)-GAN}$ , a generator and discriminator with attention maps $G_{(A+AD)-GAN}$ (proposed VAGAN) were compared. All the loss functions for the corresponding networks are summarized in Table I. In addition, BM3D [11] and $WGAN_{VGG}$ [32] were selected as two state-of-the-art post-processing techniques for comparison.

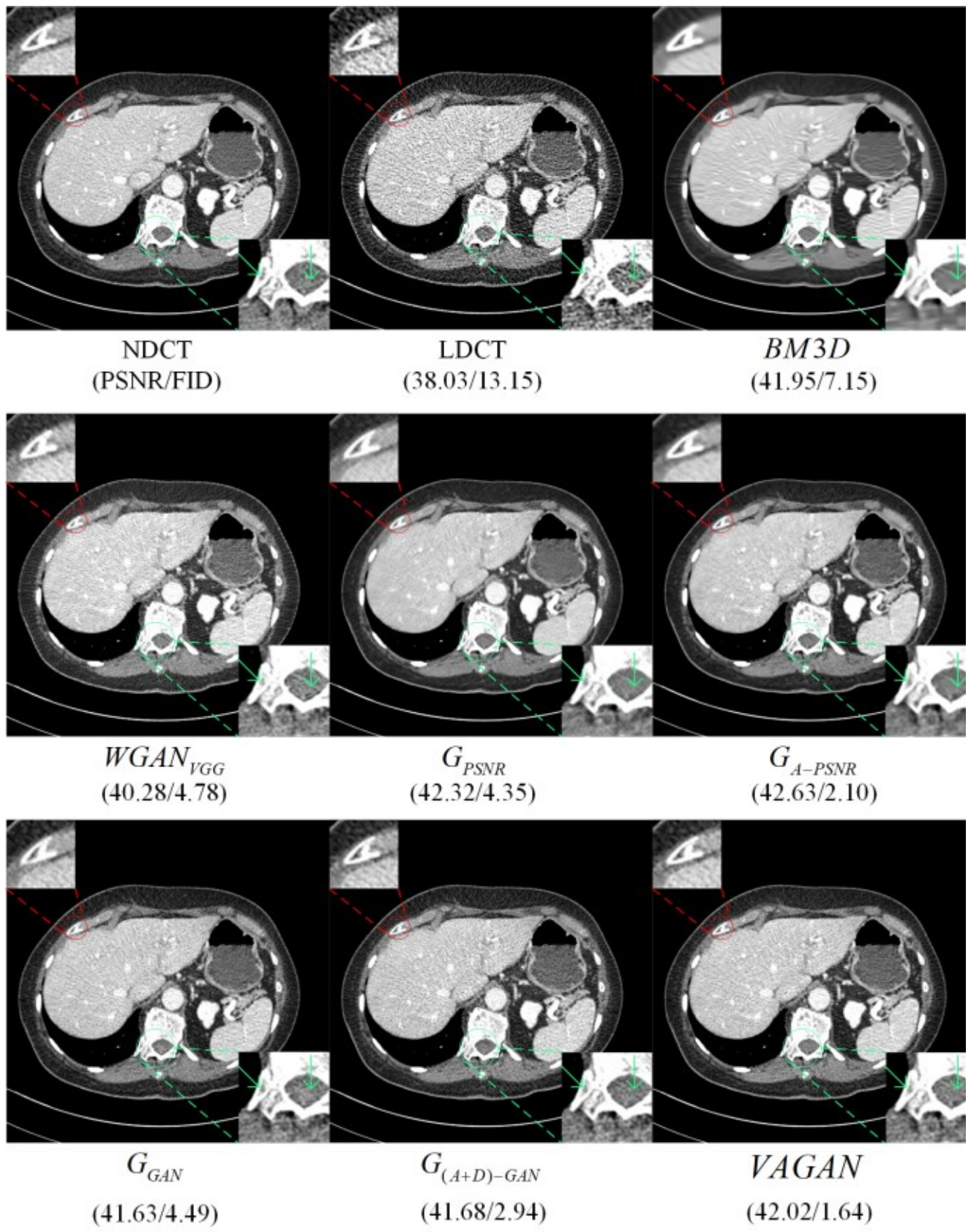

Fig. 3. Results from a transverse abdomen CT image. The display window is [-160, 240]HU.

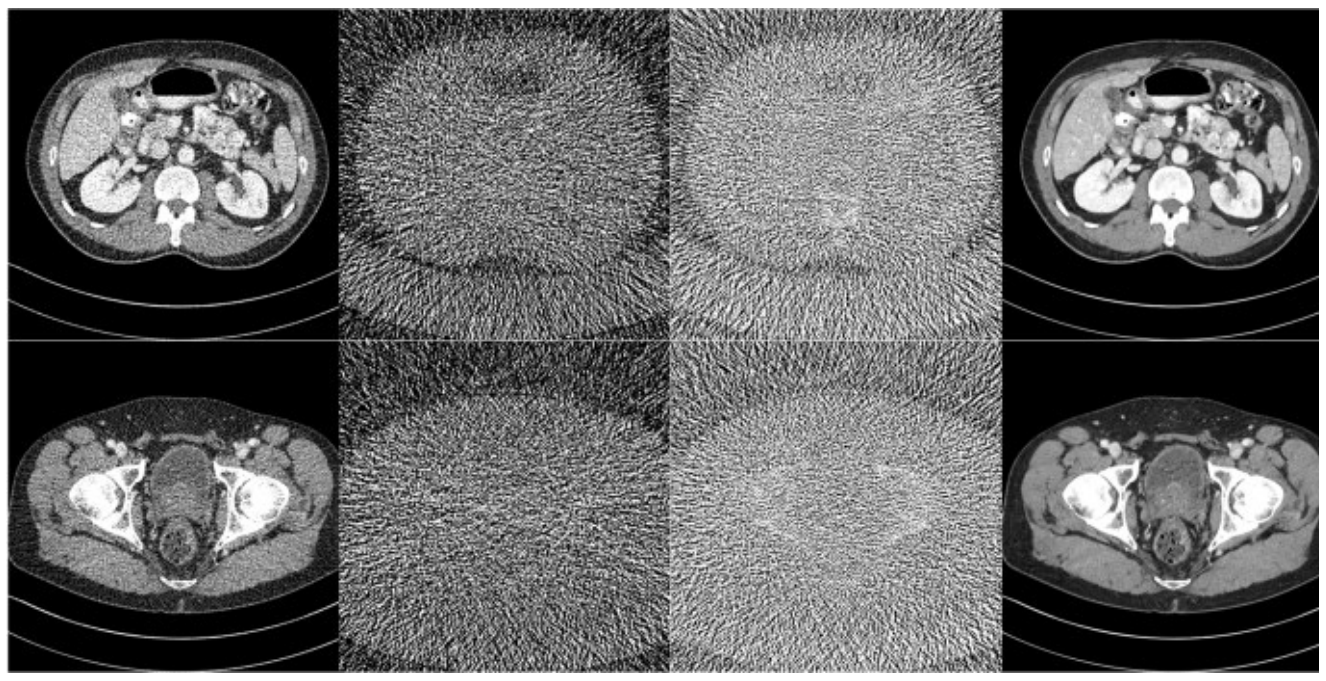

Fig. 4. Visualization of the attention maps generated by the attentive block. The first col shows LDCT images $x$ , the second col the corresponding residual maps $x'$ between NDCT and LDCT images, the third col is the attention maps $M'$ acquired by the attentive block, and the last col is denoised images $\hat{y}$ generated by PSNR-based generator.

## *C. Results*

TABLE II
QUANTITATIVE RESULTS OBTAINED BY DIFFERENT METHODS IN THE WHOLE TESTSET

| Method | PSNR | SSIM | FID |
|---|---|---|---|
| LDCT | 38.127125 | 0.960859 | 7.068731 |
| *BM3D* | 41.931743 | 0.982883 | 5.202568 |
| $WGAN_{VGG}$ | 40.367280 | 0.979806 | 1.072348 |
| $G_{PSNR}$ | 42.373726 | 0.985001 | 1.955931 |
| $G_{A-PSNR}$ | **42.807186** | **0.985612** | 1.566883 |
| $G_{GAN}$ | 41.745208 | 0.982770 | 1.114522 |
| $G_{(A+D)-GAN}$ | 41.797492 | 0.983123 | 0.964600 |
| *VAGAN* | 42.050948 | 0.985250 | **0.502748** |

Figs. 2 and 3 show the visualized results and corresponding quantitative indexes from two representative slices processed using different methods. Specially, the red circles indicate the lesion region in Fig. 2 and green circles denote some structural detail regions. It is clear that all the networks had superior abilities in image denoising/restoration. However, BM3D introduces over-smoothed effect with some waxy artifacts. $WGAN_{VGG}$ has a better visual effect but with some perceptible artifacts. Compared with $G_{PSNR}$ , $G_{A-PSNR}$ with attention map has a more powerful ability in noise removal and structural preservation, but they still tend to generate over-smoothed results. $G_{GAN}$ , $G_{(A+D)-GAN}$ and VAGAN have better visual perception and finer structural details, and the proposed VAGAN achieved best results in details restoration and perception similarity.

For further evaluate the proposed methods, quantitative results for the whole test set, including PSNR, SSIM and FID are summarized in Table II. It can be seen that VAGAN achieved competing results. In addition, the average execution time of the proposed VAGAN is about 80ms for a single slice ($512 \times 512$pix) with GPU card during the testing stage.

To demonstrate the benefits from the visual attention mechanism, Fig. 4 shows two cases with estimated attention maps. It can be observed that the generated attention maps are quite similar with the distributions of real noise, which can guide the denoising procedure efficiently.

# IV. Discussions And Conclusion

Inspired by the visual attention mechanism, in this paper we have introduced the visual attention information into the GAN framework for low-dose CT image denoising/restoration. Considering that GAN is a weakly supervised generative model, it is difficult to precisely recover corresponding NDCT images without additional information. As reported above, the results of our proposed VAGAN have been encouraging, aided by learned visual attention clues. The experimental results have demonstrated the proposed method outperforms competing methods both qualitatively and quantitatively. Compared with the other methods, our method seems superior in both noise suppression and detail preservation. Systematic evaluation and task-based optimization are in progress.